\documentclass[a4paper,10pt]{llncs}

\usepackage{graphicx}

\title{Call Graph Profiling for Multi Agent Systems}

\author{Dinh Doan Van Bien, David Lillis and Rem W. Collier}

\institute{School of Computer Science and Informatics \\ University College Dublin \\
\email{dinh@doanvanbien.com, \{david.lillis, rem.collier\}@ucd.ie}
}

\begin{document}
\bibliographystyle{splncs}
\maketitle

\begin{abstract}
The design, implementation and testing of Multi Agent Systems is typically a very complex task. While a number of specialist agent programming languages and toolkits have been created to aid in the development of such systems, the provision of associated development tools still lags behind those available for other programming paradigms. This includes tools such as debuggers and profilers to help analyse system behaviour, performance and efficiency. \emph{AgentSpotter} is a profiling tool designed specifically to operate on the concepts of agent-oriented programming. This paper extends previous work on AgentSpotter by discussing its \emph{Call Graph View}, which presents system performance information, with reference to the communication between the agents in the system. This is aimed at aiding developers in examining the effect that agent communication has on the processing requirements of the system.
\end{abstract}

\section{Introduction} \label{sect:intro}

By its nature, a Multi Agent System (MAS) is a complex system consisting of loosely-coupled autonomous software entities that are required to communicate with one another in order to achieve individual or system objectives. To facilitate the development of such systems, a number of agent-oriented programming languages and MAS toolkits have been developed by a variety of researchers~\cite{Bordini2006}. However, the availability of ancillary tools to aid with debugging and profiling is limited, particularly when compared with the available tools for other programming paradigms and languages.

Profiling is a performance analysis technique that is based on the notion that in a program, only a few places, called \emph{bottlenecks} or \emph{hot spots}, can account for the majority of the execution time of a program. Hence, by fixing only these sections of the code, the performance of a program can be substantially improved. Profiling was introduced almost 40 years ago by Donald E. Knuth in his empirical study of FORTRAN programs~\cite{Knuth71}, and has since been successfully adapted to a variety of different languages, platforms and software architectures, including large distributed systems.

\emph{AgentSpotter} is a profiling tool designed specifically for MASs. Its aim is to map traditional concepts of profiling to those of agent oriented software engineering so as to facilitate the compilation of profiling data that is specifically relevant to MAS programming. This data can then be presented in an intuitive, visual fashion in order to aid multi agent developers in improving the performance of their systems. Previous work outlined the AgentSpotter architecture and its Space-Time Diagram view, which was aimed at identifying events in a system that impact performance~\cite{dvb2009}. 

This paper continues this work by introducing AgentSpotter's Call Graph View, which attempts to contextualise performance information by linking it to causal events, such as message passing.

Section~\ref{sect:related} provides a brief discussion of some related tools that have been developed for debugging and profiling MASs. In Section~\ref{sect:agentspotter}, we give a brief overview of the AgentSpotter agent profiling application. Following this, in Section~\ref{sect:call_graph} we introduce the concept of a \emph{call graph}, and analyse how the traditional concept of a call graph can be applied to a MAS. Section~\ref{sect:visualisation} presents the concrete implementation of an agent call graph within the AgentSpotter profiling tool, followed by a discussion of the proposed approach in Section~\ref{sect:discussion}. Finally, we conclude and outline ideas for future work in Section~\ref{sect:conclusions}.

\section{Related Work} \label{sect:related}

The work presented in this paper draws from two principal research areas. Firstly, in order to provide a profiling tool for MASs, it is necessary to examine the concepts and features of existing profiling tools for other programming paradigms, such as object-oriented programming. It is also necessary to explore the available programming tools aimed at aiding the debugging and profiling of MASs.

Initially proposed by Knuth, the key motivating factor behind profiling tools is his observation that ``less than 4\% of a program accounts for more than half of its running time''~\cite{Knuth71}. By identifying and improving code that represents a performance bottleneck, software developers can greatly improve the overall performance of their programs. An important motivator for the use of specialist profilers to identify these bottlenecks is the frequent tendency of developers' mental maps of their programming not to match the reality of how their programs behave. Thus, a profiler may identify areas of concern that a programmer may not have considered.

In the context of more traditional, non-MAS, programming, developers generally have access to long-established and widely-accepted profiling tools such as gprof~\cite{Graham82gprof} or performance analysis APIs such as the Java Virtual Machine Tool Interface (JVMTI)~\cite{jvmti2004} or ATOM~\cite{Srivastava1994}. However, those developing MASs do not tend to have access to such well-established tools.

One MAS framework that does provide the ability to glean data about system performance is Cougaar~\cite{Helsinger2004}. This provides access to data on historical performance data, event detection, monitoring of ACL messages and a number of other services. The LS/TS agent platform provides an administrator tool that records some high-level system monitoring information~\cite{Rimassa2005}. The main limitation of these systems is the lack of post-processing of the raw performance data in order to produce meaningful synthetic indicators like a profiler would do.

Besides performance analysis, most agent frameworks provide a debugging tool similar to the Agent Factory Debugger~\cite{Collier2007}, which provides information about the mental state and communication from the viewpoint of individual agents. A different type of debugging tool is the Agent Viewer that is provided in the Brahms toolkit~\cite{Seah2005}, which displays agent timelines so as to understand when agents' actions are taken.

As the work in this paper also requires the monitoring of inter-agent communication (see Section~\ref{sect:call_graph}), it is also important to acknowledge the availability of existing communication analysis tools for MAS platforms. A number of such tools have been developed for a variety of agent frameworks and toolkits to aid developers in understanding the interaction between agents in their systems. An early example of such a toolkit is Zeus~\cite{Nwana1998}, which contains a ``society tool'' that visualises the interaction between agents, so as to help in understanding the topology of the social contacts within the MAS. This type of tool also aids in debugging MASs, since developers can ensure that the expected communication and collaboration between agents is indeed taking place.

In the JADE agent development framework, a Sniffer Agent is a FIPA-compliant agent that monitors messages created with an Agent Communication Language (ACL) passed between agents and presents these in a simple graphical interface~\cite{Bellifemine2007}. A more sophisticated tool, called \emph{ACLAnalyser}, provides more detailed information on agent communication~\cite{Botia2005}. Again, the principal aim of this is to aid in debugging errors in MASs that relate to coordination or cooperation.

\section{AgentSpotter} \label{sect:agentspotter}

\begin{figure}[!h]
    \centering
    \includegraphics[scale=0.45]{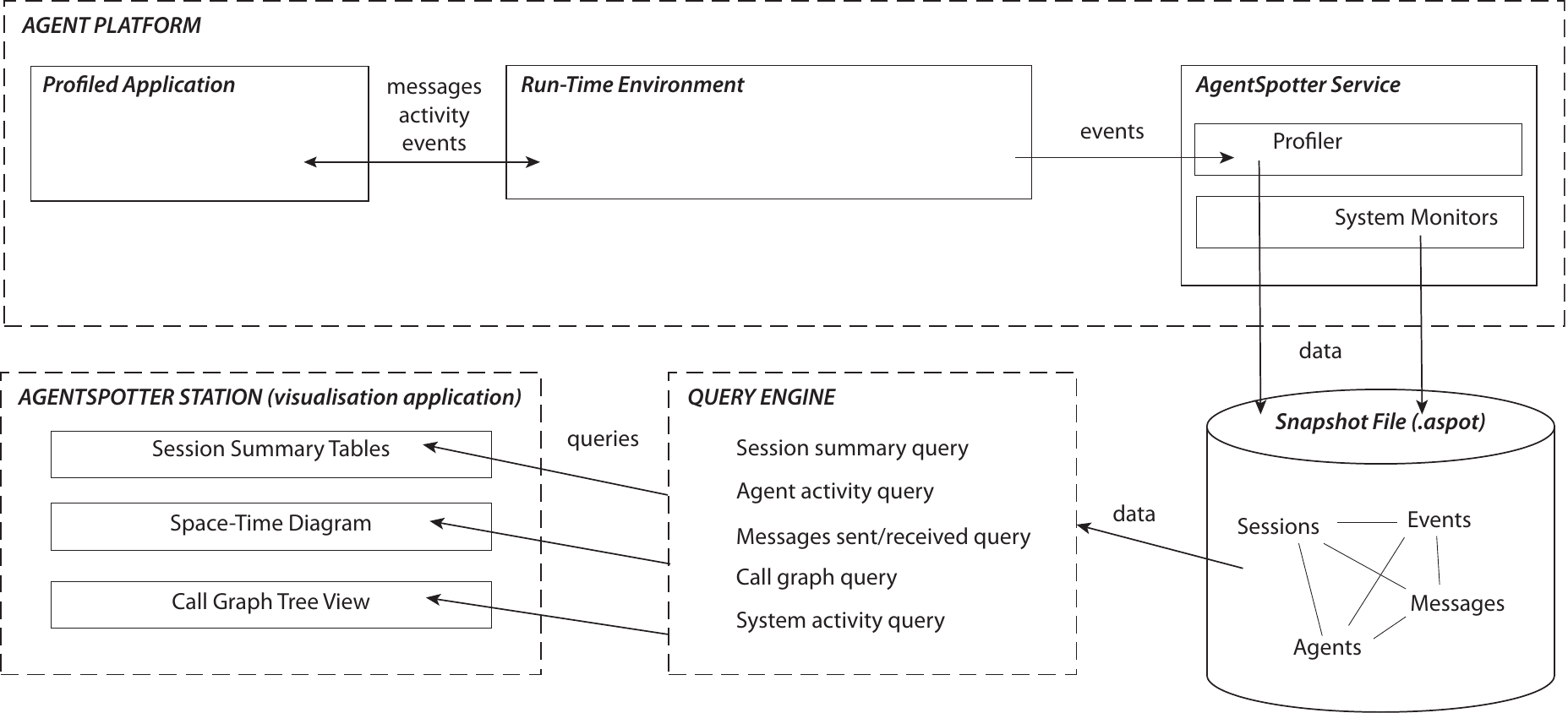}
    \caption{AgentSpotter Architecture}
    \label{fig:abstract_architecture}  
\end{figure}

AgentSpotter is a profiling tool designed specifically for gathering and displaying profiling information on MASs~\cite{dvb2009}. Figure~\ref{fig:abstract_architecture} illustrates the abstract architecture of the system, designed to maximise compatibility with any type of agent platform. The \emph{AgentSpotter Service} runs within the Run-Time Environment of an Agent Platform, gathering data about the agents themselves (actions performed, messages exchanged), along with system data such as CPU and memory usage. This is the only platform-specific portion of the system that must be ported in order to allow AgentSpotter to be run on different agent platforms. The data gathered is logged into a Snapshot File, which allows it to be accessed and analysed offline, once the system has finished running.

AgentSpotter Station is a visual application that provides a number of visualisations on various aspects of system performance, in order to help programmers to identify performance bottlenecks in their code.

The outputs utilised in this paper are gleaned from running a concrete implementation of the AgentSpotter service within the Agent Factory framework~\cite{Collier2003}. Agent Factory is a modular and extensible framework that provides comprehensive support for the development and deployment of agent-oriented applications. A more detailed description of this implementation and the data gathered by AgentSpotter can be found in~\cite{dvb2009}.

Previously, it was shown how AgentSpotter was used to map traditional profiling concepts on to agent-oriented concepts. This focused on two types of visualisation:
\begin{itemize}
	\item \textbf{Flat Profile:} provides data on such things as agent activity, messages and reasoning/action duration in a tabular form.
	\item \textbf{Space-Time Diagram:} provides a navigable visualisation representing the data from the flat profile in a more intuitive manner.
\end{itemize}

The focus of this paper is on an agent-oriented call graph. Whereas a space-time diagram can aid in identifying the timing and extent of actions executed by agents, a call-graph is traditionally intended to also indicate the reasons why particular actions were undertaken at particular times.

\section{Call Graph Concept}
\label{sect:call_graph}

\subsection{Traditional Call Graph}

The concept of a call graph was introduced in 1982 in the ``gprof'' profiling tool~\cite{Graham82gprof}. This is an improvement on the popular ``prof'' UNIX profiling tool. In additional to summarising the time spent in different functions, it also recursively presents all the call stacks annotated with the time spent in the various functions that are called. Another name for the call graph is ``hierarchical profile'', which conveys the idea that gprof provides information to aid in understanding the impact of one function in relation to all the functions that have called it.

Although the textual output of gprof is very dense and requires some practice to understand, the user interfaces of modern profiler have made call graphs more tractable by presenting them as tree view controls that can be interactively explored.

\begin{figure}[!h]
    \centering
    \includegraphics[scale=0.4]{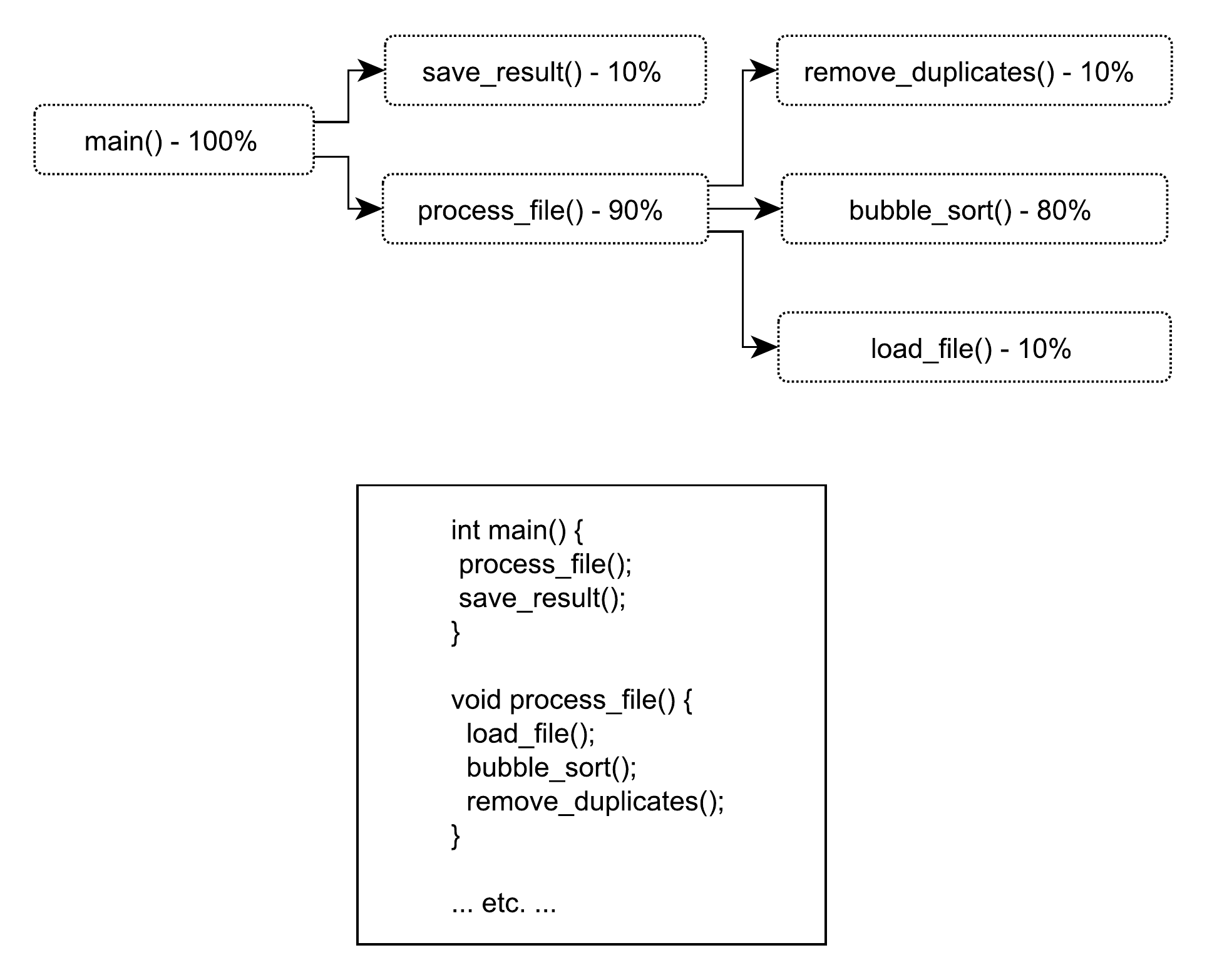}
    \caption{Call Graph Tree View of a fictional C program that removes duplicate lines from a file}
    \label{fig:callgraph_traditional}  
\end{figure}

Figure~\ref{fig:callgraph_traditional} shows an example of a typical call graph. This relates to a fictional C language program that is designed to remove duplicate lines from a text file. In this tree view, the root node is the \emph{main} function, which represents 100\% of the total execution time of the program (including the execution time of its child nodes). Each node represents a function within the program, with the child nodes representing functions that are called within the parent function. The percentages represent the cumulative proportion of the program's execution time that is attributable to a node and its children. In the example, the \emph{main} function calls \emph{process\_file}, which is then represented as a sub-tree with leaves representing its own calls to the \emph{bubble\_sort}, \emph{load\_file} and \emph{remove\_duplicates} functions.

The key benefit of the call graph tree view is the extended context it gives to performance information. For instance, this simple example reveals that the program spends 90\% of its time processing a file. The tree shows that one of the top-level function \emph{process\_file}'s callees, the \emph{bubble\_sort} operation, accounts for 80\% of its caller time. A flat profile would have shown the time for these functions separately without explicitly showing the hierarchical link between them.

\subsection{Agent-Oriented Call Graph Model}
\label{sect:call_graph_model}

When constructing a flat profile for a MAS, it was necessary to map a number of concepts relating to traditional programming to equivalent concepts in the domain of agent-oriented programming~\cite{dvb2009}. A similar mapping must be performed in order to allow for the development of an agent-oriented call graph.

The central measure used in the traditional call graph is the function execution time. Each node represents a function, which can take the action of calling other functions as part of its execution. The consequence of this action is that some amount of time is spent executing the child function. Thus we can say that the \emph{impact} of calling a function is that this additional processing time has been incurred.

In many MASs, agents tend to perform actions as a reaction to the receipt of ACL messages from other agents in the system. Thus in the same way the impact of a functional call in a traditional system is the execution time of that function, within a MAS, the impact of a message can be related to the additional processing that must be undertaken in order to react to the information contained therein, formulate a response or perform a requested task. Because of this mapping, we introduce, as a first simplified approach, the \emph{agent message impact} measure to be used as an equivalent to the function processing time used in traditional profiling. It is important to acknowledge that not all MASs are reactive in nature and other events (such as those arising from an agent's environment) that are not represented by this measure. Further discussion can be found in Section~\ref{sect:discussion}.

The quantification of such a measure is a difficult task, given the data typically available from MASs. One potential measurement for $T_{{M_\alpha} , B}$, the impact of a message $M_\alpha$ sent from an agent $A$ to an agent $B$ and received at time stamp $\alpha$ is to use the total amount of computation time used by the agent $B$ until agent $B$ receives a message $M_\Omega$ from another agent $X$ at time stamp $\Omega \geq \alpha$. Let $b$ be the duration of an activity by agent $B$ at time stamp $t$ where $\alpha \leq t \leq \Omega$. The impact of message $M_\alpha$ on agent $B$, $T_{{M_\alpha} , B}$, is then given by the recurrent equation:
\begin{equation}
T_{{M_\alpha} , B} = \displaystyle\sum_{t=\alpha}^{\Omega} b_t
\label{equation-message_impact}
\end{equation}

\begin{figure}[!h]
    \centering
    \includegraphics[scale=0.65]{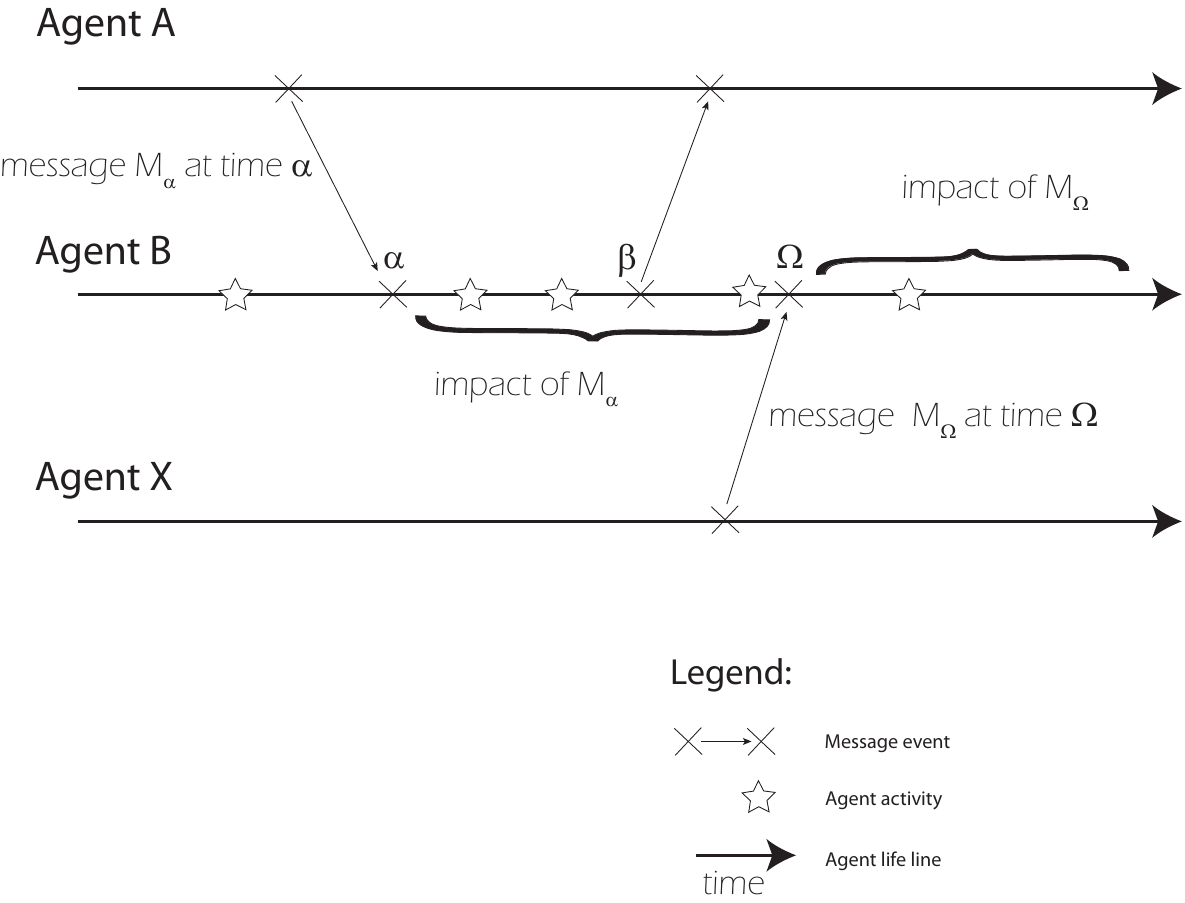}
    \caption{Agent Message Impact Concept Diagram}
    \label{fig:agent_message_impact}  
\end{figure}

In Figure~\ref{fig:agent_message_impact} we have tried to summarise this concept in a graphical form. The diagram clearly shows that the three activity stars that lie between $\alpha$ and $\Omega$ make up the total impact of $M_\alpha$ on agent $B$. Note that the outgoing message at time stamp $\beta$ does not break the computation sequence.

It is now easy to determine the total impact $T_{x , y}$ of all the messages sent by a given agent $x$ to another agent $y$. Let $M$ be the total number of messages sent, $1 \leq m \leq M$ a single message impact identifier, $\alpha_m$ the reception time stamp of message $m$ from $x$ to $y$, and $\Omega_m$, where $\alpha_m \leq \Omega_m$, the next reception time stamp message coming right after $m$ from any other source. The total impact $T_{x , y}$ is then given by the equation:

\begin{equation}
T_{x , y} =
	\displaystyle\sum_{m = 1}^{M}
	\displaystyle\sum_{t=\alpha_m}^{\Omega_m} b_t
\label{equation-agent_x_y_impact}
\end{equation}

By applying the equations recursively, we can compute the total impact $T_x$ of an agent $x$ on $N$ other agents numbered $1 \leq a \leq N$ as follows:

\begin{equation}
T_{x} =
	\displaystyle\sum_{a = 1}^{N}
	\displaystyle\sum_{m = 1}^{M_a}
	\displaystyle\sum_{t=\alpha_m}^{\Omega_m} b_t
\label{equation-agent_x_impact}
\end{equation}

Finally, the total impact $T_S$ of all the $K$ agents numbered $1 \leq k \leq K$  of a session $S$ is given by the equation:

\begin{equation}
T_{S} =
    \displaystyle\sum_{k = 1}^{K}
	\displaystyle\sum_{a = 1}^{N_k}
	\displaystyle\sum_{m = 1}^{M_a}
	\displaystyle\sum_{t=\alpha_m}^{\Omega_m} b_t
\label{equation-total_session_impact}
\end{equation}

It must be noted that the total activity time $A_S$ of the session $S$ is given by the equation:

\begin{equation}
A_{S} = T_S + 
    \displaystyle\sum_{k = 1}^{K}
	\displaystyle\sum_{t=\alpha_S}^{\alpha_{k0} - 1} b_t
\label{equation-total_session_activity}
\end{equation}

where $\alpha_S$ is the first recorded time stamp in session $S$ and $\alpha_{k0}$ the time stamp of the very first message received by agent $k$. To put it differently, the total impact for each agent can be computed only after it has received its first message.

This proposed method of calculating agent message impact is imperfect, and superior metrics are likely to be developed in the future. However, it does provide useful information for the debugging and development of MASs. Both the drawbacks and benefits of this approach are outlined in more detail in Section~\ref{sect:discussion}.

\section{Call Graph Visualisation Specification} \label{sect:visualisation}

\begin{figure}[h]
    \centering
    \includegraphics[scale=0.45]{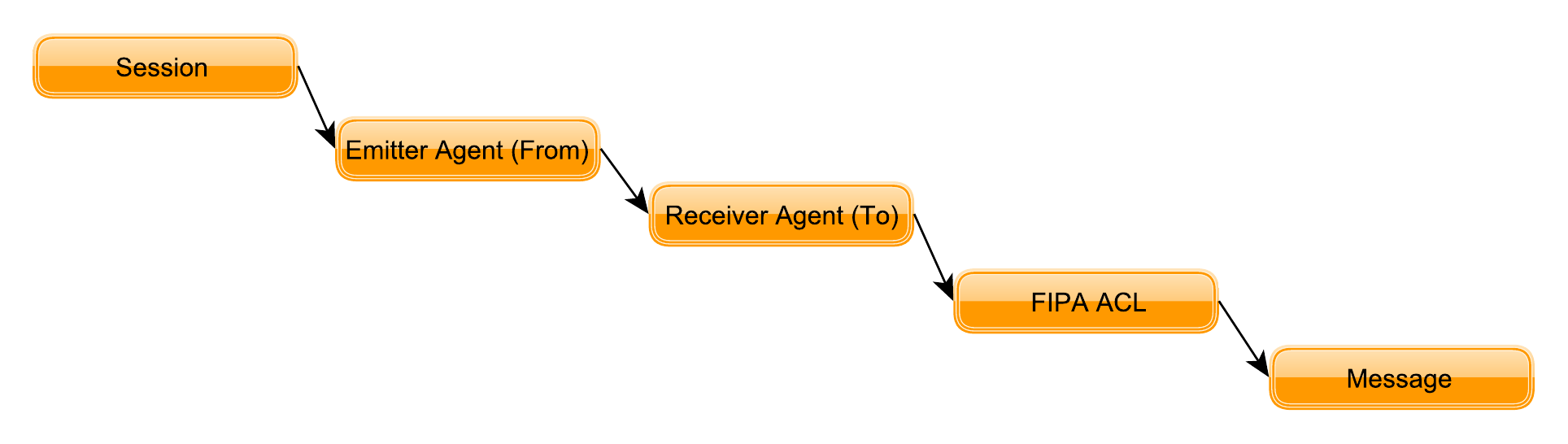}
    \caption{Call Graph Tree View levels}
    \label{figure-call_graph_levels}  
\end{figure}

The conceptual model we have presented deals with the session level, the emitter agent level, the receiver agent level and the message level. The graphical translation of the model, outlined In Figure~\ref{figure-call_graph_levels}, should be a tree view representing the levels we have previously enumerated plus an additional level for the FIPA ACL message content. A message content is defined as a performative plus an expression e.g. ``request:doSomeThing(123)''. This additional level should give developers necessary contextual information for the messages. It is important to note that this fixed-depth call graph tree represents a divergence from traditional call graphs, whose depth is dictated by the depth of the deepest function call stack.

The session at the root of the tree should add up to 100\% of all emitter agents' impact as defined by Equation~\ref{equation-total_session_impact}. Then at each level, each node should recursively total the impact of its child nodes down to the message leaf nodes. These leaf nodes simply report their impact as defined by Equation~\ref{equation-message_impact}. More precisely, at each level, for each node, the following values should be displayed:
\begin{itemize}
\item \textbf{Label:} informative text associated with the node. The structure of the label depends on the level as follows:
\begin{itemize}
\item session: ``capture date, time - duration'';
\item emitter agent: ``from: agent id'';
\item sender agent: ``to: agent id'';
\item FIPA ACL: ``performative: contents'';
\item message: ``sent: time stamp rec: time stamp''.
\end{itemize}
\item \textbf{Total impact time:} sum of impact times of all the current node's children.
\item \textbf{\% parent time:} percentage of the current node total impact time divided by the node's parent total impact time.
\item \textbf{\% session time:} percentage of the current node total impact time divided by the session total impact time.
\end{itemize}

Ideally, developers should be able to order the intermediary tree levels differently so as to produce different call graph interpretations. For example, moving the FIPA level right above the emitter agent level would list for each FIPA ACL entry their total impact for all the emitter/receiver pairs.

\subsection{User Interface} \label{sect:ui}

Despite having a fixed depth, a call graph tree view could potentially be very wide at the leaf level for sessions that produce thousands of messages. Therefore, to help developers navigate easily through the tree, AgentSpotter Station offers an advanced tree navigation user interface that expands only that part of the tree that is currently explored so as to reduce the visual clutter. The currently explored part of the tree is highlighted in a different colour to give the developer some visual feedback. 

Moreover, to speed up the retrieval of information on the system, a search feature allows developers to enter a keyword (e.g. an agent name or a performative). Doing so has the effect of highlighting in a special colour all the visible nodes that contain the specified keyword, significantly improving the visual retrieval speed of a node.

Finally, developers can zoom and pan around the tree view to locate items even more quickly.

\subsection{Implementation}

\begin{figure*}[htp]
    \centering    
    \includegraphics[scale=0.43]{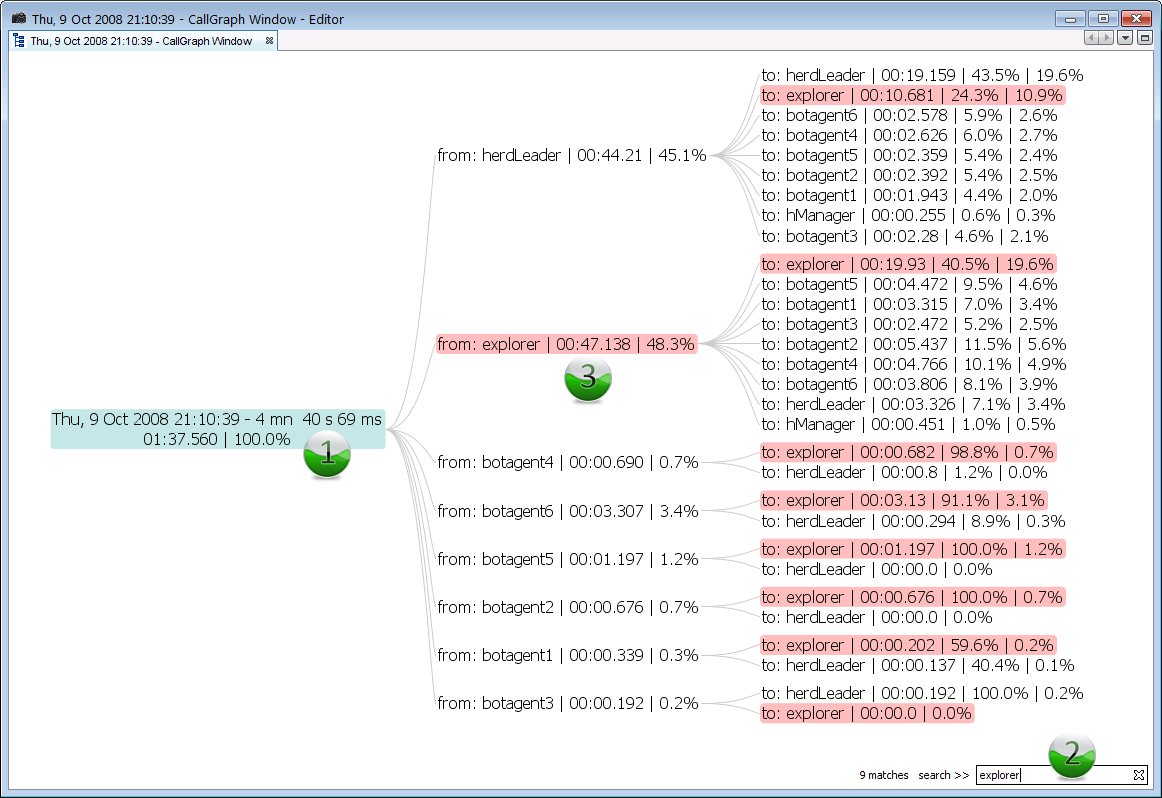}
    \caption{Call Graph Tree View Screen Shot}
    \label{fig:callgraph_snap_01}  
\end{figure*}

A sample screen shot of the visualisation of the call graph can be seen in Figure~\ref{fig:callgraph_snap_01}.

In this figure, the element numbered 1 on the screen shot is the tree root, i.e. the session level which represents 100\% of the cumulative recorded activity time. The tree root is highlighted in blue because it is the current tree selection in this specific example. As such, it determines the branch that is expanded, as stated in  Section~\ref{sect:ui}, so as to reduce the visual clutter. In order to provide a sufficient level of detail, all the children  and grandchildren of a selected node are visible. Consequently, when the tree root is selected, only the first two subsequent levels are expanded, that is the emitter agent level and the receiver agent level. Hence, selecting an emitter agent node should make the FIPA ACL message level visible, and so on. As an illustration, the call graph numbered 3 shown in Figure~\ref{fig:callgraph_snap_01} screen shot, has an agent receiver node selected; as a result, this branch is fully expanded down to the message impact level.

The element numbered 2 is a text area used to enter a search keyword. The number of nodes matching the keyword is displayed and all the matching nodes that are visible are highlighted in pink. For instance, the element numbered 3 is one of the nine nodes containing the ``explorer'' keyword and so is highlighted in the screenshot.  In a large expanded tree, this highlighting greatly adds to the visual effect and consequently to the navigability of the tree. The bottom-most highlighted node in the tree represents a message sent from the ``botagent3'' agent to the agent named ``explorer''. Clicking on this node would cause the subtree rooted at that node to be expanded so as to examine the content and timing of that message.

The visualisation is completely interactive and  can be controlled using the mouse or the keyboard. Possible interactions include panning, scrolling, expand tree branches, zooming in and out.

One other important feature is the ability to alter the hierarchy of the nodes. Whereas the recursive nature of function calls means that these are inherently inflexible in the tree hierarchy they create, the nature of message-passing is a different situation. The hierarchy above places the sender of each message in a higher position in the hierarchy than the recipient. This means that the cumulative performance data for the higher-level nodes represents the overall impact of all messages sent by a particular agent to other agents. However, this may not encapsulate the information that a developer requires at a particular point in time. Changing the hierarchy to place the recipient agent above the sender changes the focus of the cumulative performance data. In this case, the figures represent the contribution to overall running time of a particular agent, based on the message that it receives from any and all sources. This may potentially identify entire individual agents as bottlenecks. This may be because the system's load is imbalanced, meaning that one agent may bear an inequitable share of the processing burden. Alternatively, in a distributed MAS, an agent may simply reside on a machine with inferior hardware resources. By exploiting the flexible nature of this hierarchy, users of the call graph tree view can alter the data being presented to better fit their needs.

\begin{figure*}[!t]
    \centering
    \includegraphics[scale=0.26]{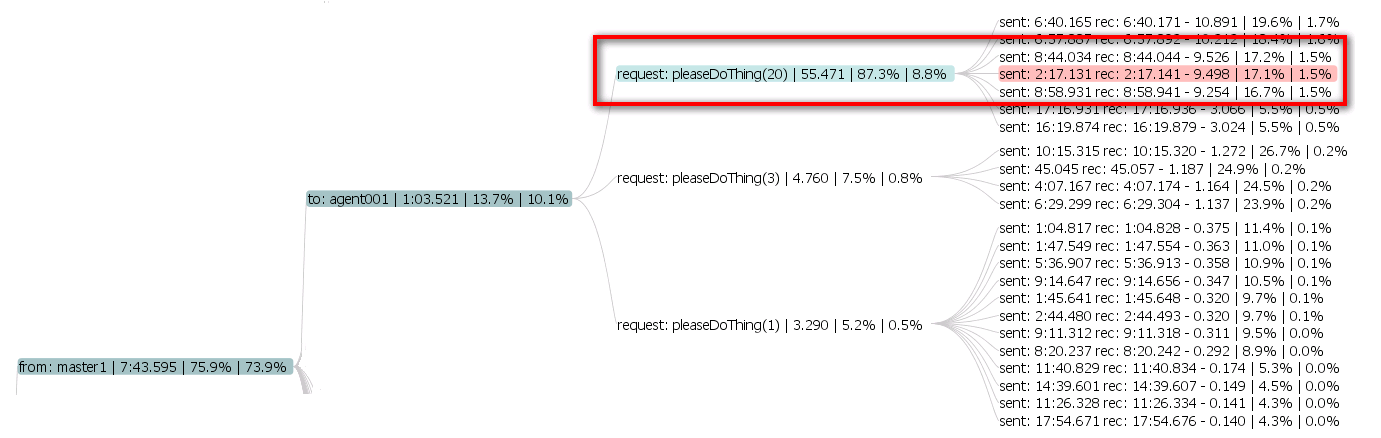}
    \includegraphics[scale=0.4]{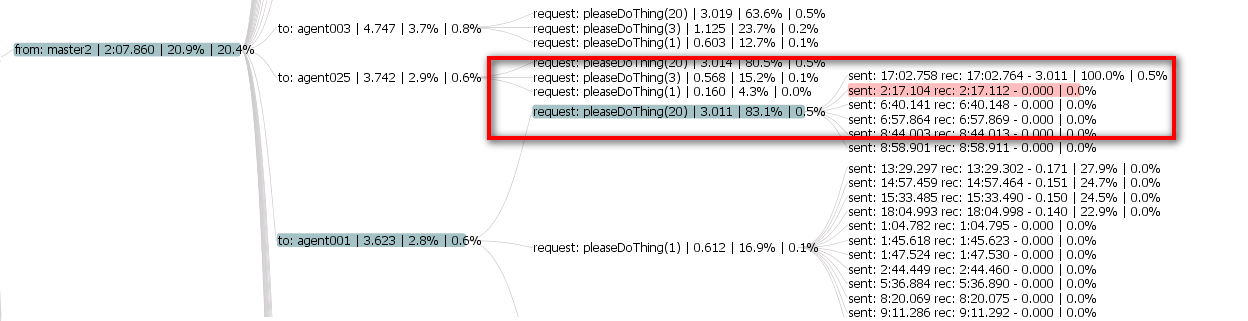}
    \caption{Benchmark Call Graph Tree View for master1 and master2}
    \label{fig:benchmark_cg_master1}
\end{figure*}

\section{Evaluation} \label{sect:evaluation}

To demonstrate the effectiveness of the call graph as a profiling tool, a simple benchmark application was developed. This consists of two types of agents. Overseer agents request worker agents to perform small, medium or large tasks. If a worker agent has recently been overloaded, it may refuse to execute the required task. Occasionally, overseer agents will delegate the assignment of tasks to a worker agent, in which case the worker agent becomes an overseer agent for a brief period. A flat profile and space-time diagram for this benchmark system is contained in~\cite{dvb2009}. Figure~\ref{fig:benchmark_cg_master1} shows a portion of the call graph tree view for a run of this application. Here, the names of overseer agents begin with ``master'', whereas the names of worker agents begin with ``agent''.

The benchmark application profile (displayed in Figure~\ref{fig:benchmark_cg_master1} reveals that overseer agents master1 and master2 do not have the same impact on performance. Intuitively, one would expect each overseer agent to have an equal impact. However, in reality, we can see that the impact of messages sent by the ``master2'' agent accounts for only 20.4\% of the overall session running time. Studying the call graph in more details helps in explaining this imbalance, by studying the effects of the messages with the content ``pleaseDoThing(20)'' that were sent by both master1 and master2 to agent001. These are emphasised in Figure~\ref{fig:benchmark_cg_master1} by means of the red rectangles. In each case, the parameter passed as part of the a ``pleaseDoThing'' request is related to the amount of work that the agent is being requested to perform.

 The call graph shows that some requests from master2 have a 0.0 impact which in practice means they were ignored (no actions took place as a result of receiving those messages). In other words, when master1 sends a request to an agent, and immediately afterwards that master2 sends the same request to the agent, the overloaded agent simply refuses to execute the request. These ``pleaseDoThing(20)'' messages sent by master1 are reasonably consistent in terms of their impact, are never refused and account for a total of 8.8\% of the total session running time. In contrast, only a single such request sent by master2 was honoured by agent001. This action accounted for a mere 0.5\% of the session running time.

It is important to note that the greater impact of master1's messages does not necessarily constitute a bottleneck, merely an imbalance in the system. This type of analysis would motivate the use of the space-time diagram to examine the timing of the messages in question, so as to further find why messages from master2 are more likely to be ignored by the worker agent.

A bottleneck would be identified by comparing the impact of different messages being sent (rather than the same message being sent by different agents). For instance, it is notable that the session impact percentages for ``pleaseDoThing(1)'' messages sent by master1 to agent001 are far lower than for ``pleaseDoThing(20)''. In this simple benchmark application, this is an unsurprising result, as the increased workload is explained by the messages themselves, with the latter message requesting more processing to be undertaken by the former. However, figures such as these would indicate a bottleneck if the results are unexpected (i.e. where high-impact messages are not intended to trigger high-cost actions on the part of the message recipients) and so would motivate a closer examination of the longer-running actions to increase efficiency.

 It may be possible to make such a deduction from viewing the underlying agent code itself, however the use of the call graph makes this far more easily apparent without the need for detailed examination of the code. This also means that testers that are not necessarily familiar with the code (or even perhaps testers who do not understand the programming language used) can identify bottlenecks and behavioural anomalies for developers to address.

\section{Discussion} \label{sect:discussion}

The proposed metric for measuring the agent message impact outlined in Section~\ref{sect:call_graph_model} has a number of drawbacks. It operates on the naive assumption that the actions of an agent are directly related to the messages received by it. The impact of a message on an agent is thus taken as the sum of the execution times of all actions undertaken by the agent between the receipt of that message and the receipt of the next message. This means that the proposed metric may not be appropriate to certain types of MAS, where agent action is not intended to occur as a result of communication with other agents.

The principal drawback with such an approach is that there is no provable causal link between the receipt of messages and the execution of actions. Agents may decide to act for reasons other than the receipt of ACL messages. For instance, a perceptor may have detected changes in the environment that may require some reaction. Also, when actions are executed as a direct consequence of the receipt of an ACL communication, there is no guarantee that all of the relevant actions have been performed prior to the receipt of the next message. Thus the agent message impact arising from the receipt of a single message may not be particularly informative.

Ideally, the best method of measuring the impact of the receipt of an ACL message would be to track the internal reasoning process of the agent, so as to identify those actions that are performed as a direct result of the receipt of a message and take only these into account when calculating the message impact. This is, however, a particularly difficult task, as the reasoning used by agents is extremely platform-dependent and would require a substantial amount of work to be performed in order to port AgentSpotter to other agent platforms and frameworks. This contravenes one of the fundamental aims of AgentSpotter, which is intended to be as platform-agnostic as is practicable.

A further difficulty arises when considering the fact that the receipt of messages is not the only event that may cause an agent to undertake some processing. An agent may also perceive events occurring in its environment, which may prompt some action being taken. Again, the mechanism by which agents perceive such events will vary between agent platforms. Furthermore, toolkits such as CArtAgO~\cite{ricci2007cartago} or SoSAA~\cite{Dragone2009}, which can be used for environment abstraction, will raise events in different ways, thus adding to the complexity and framework-dependence of the task. In contrast, FIPA-compliant ACL messages will be standardised across toolkits and platforms.

Even if we are to settle for a framework-specific profiling system, the task of identifying direct causal links between events is also non-trivial. Whereas some agents may contain straightforward agent code that reacts to the receipt of a message or an environmental percept by always invoking a particular action, this is unlikely to always be the case. Events may instead lead to a refinement of an agent's goals, or even more subtly, an alteration of its current belief state, which in turn may result in goal refinements. Goals may be adopted based on the entire belief set, making it difficult to ascertain for certain whether the belief triggered by the message was a cause of the change in the agent's goals or a merely coincidental occurrence. Even when goals have been adopted, a plan selection algorithm is typically used to decide upon the best path to take towards satisfying those goals. Again, this is a potentially difficult process to trace reliably.

Although the proposed measure does have drawbacks and is somewhat simplistic, it is also important to highlight the benefits of such a measure. Whereas a high impact measurement for a single message may not be indicative of a major system bottleneck (and may indeed be merely coincidental), consistently high impact measures for similar types of messages are far more likely to be a result of a causal link between the receipt of the message and the processing that follows. Thus by providing an approximation of the communication dynamics within a MAS, valuable insights may be gained.

It is this type of analysis that makes the call graph a useful tool in identifying situations that result in a high processing load and thus aid in helping developers concentrate on the appropriate portions of the code base to improve system efficiency. 

\section{Conclusions and Future Work} \label{sect:conclusions}

We have proposed a new visualisation, the call graph tree view, in order to provide detailed information about the performance impact of agents interactions. After discussing the concept of call graph in a traditional programming context, we have then mapped it into an agent-oriented concept based on the idea that when an agent sends an ACL message to another, its impact on the amount of processing the recipient perform can be measured and used to identify system bottlenecks, load imbalances and efficiency issues. Although the proposed measure is not optimal, it does provide users with data that is appropriate and useful in the context of a profiler application. We have then extended this notion to a tree model with multiple levels: session, message emitter, message receiver, message. Finally, we have described the advanced user interface that allows developers and testers to interact with this model in the form of a zoomable and searchable tree view.

For further development of the call graph view of the AgentSpotter application, there are two principal areas for improvement. Firstly, as we have acknowledged in Section~\ref{sect:discussion}, the current measure for gauging the processing impact of a message being passed between agents is not an ideal one. We intend to investigate other possible measures that will include a stronger causal link between the receipt of a message and the resulting processing activity. In doing so, the other AgentSpotter views (flat profile and space-time diagram) will be utilised to ensure that any proposed measures reflect the reality of the system's execution as closely as possible.

The second significant area of future work is in the area of agent conversation protocols. The work presented in this paper considers each ACL message to be entirely independent of all other messages. The reality of agent communication is somewhat different. In the agent architecture presented as the benchmark application in Section~\ref{sect:evaluation}, some overseer agents request that worker agents perform certain tasks. In our simple application, this is done by means of a single message containing the work request being sent to the worker. In reality, a more complex conversation would be used. The initial request for a task to be performed may be answered with an acceptance or rejection of the task being assigned, followed perhaps by the communication of the result of the task. Clearly, an agent accepting and performing a task will consume more processing resources than when it is rejected. However, in the existing model, both scenarios will be grouped together, under the initial message requesting action. Such behaviour may mask inefficiencies in the processing code by including the low-cost rejection actions in its session percentages. By introducing an additional conversation level into the tree, these situations can be separated, meaning that actions will be grouped according to entire agent transactions rather than single messages.

\bibliography{thesis,spe,promas,master}

\end{document}